 \documentclass[journal, comsoc]{IEEEtran}

\usepackage{gensymb}
\usepackage{cite}
\usepackage{amsmath,amssymb,amsfonts}
\usepackage{graphicx}
\usepackage{xcolor}
\usepackage{kotex}  

\usepackage{caption}
\usepackage{subcaption}

\usepackage{stfloats}
\usepackage{float}
\usepackage{wrapfig}
\usepackage{xcolor}
\usepackage{colortbl} 
\usepackage{color,soul}
\usepackage{url}
\usepackage{verbatim}
\usepackage{graphicx}
\usepackage{cite}

\usepackage{multirow}
\usepackage{array,boldline, makecell,booktabs}
\usepackage{longtable} 

    \usepackage[group-separator={,},group-minimum-digits=4]{siunitx}

\begin{document}
%
\title{Towards Integrated Sensing and Communications for 6G: A Standardization Perspective}
%
%
%

\author{Aryan Kaushik, Rohit Singh, Shalanika Dayarathna, Rajitha Senanayake‬, Marco Di Renzo, Miguel Dajer, Hyoungju Ji, Younsun Kim, Vincenzo Sciancalepore, Alessio Zappone, and  Wonjae Shin\thanks{A. Kaushik is with the School of Engineering \& Informatics, University	of Sussex, UK (e-mail: aryan.kaushik@sussex.ac.uk). \\$~~~$R. Singh is with the Department of Electronics \& Communication Engineering, Dr. B. R. Ambedkar National Institute of Technology, India (e-mail: rohits@nitj.ac.in). \\$~~~$S. Dayarathna and R. Senanayake are with the Department of Electrical \& Electronic Engineering, University of Melbourne, Australia (e-mails: \{s.dayarathna, rajitha.senanayake\}@unimelb.edu.au). \\$~~~$M. Di Renzo is with the CentraleSupelec, Paris-Saclay University, France (e-mail: marco.di-renzo@universite-paris-saclay.fr). \\$~~~$M. Dajer is with the Futurewei Technologies, USA (e-mail: mdajer@futurewei.com). \\$~~~$H. Ji and Y. Kim are with Samsung Research, Samsung Electronics, South Korea (e-mails: \{hyoungju.ji, younsun\}@samsung.com).\\$~~~$V. Sciancalepore, is with the NEC Laboratories Europe GmbH, Germany (e-mail: vincenzo.sciancalepore@neclab.eu). \\$~~~$A. Zappone is with the Department of Electrical and Information Engineering, University of Cassino and Southern Lazio, Italy (e-mail: \{alessio.zappone@unicas.it).  \\ $~~~$W. Shin is with the Department of Electrical and Computer Engineering, Ajou University, South Korea, and also with the Department of Electrical Engineering, Princeton University, USA (e-mail: wjshin@ajou.ac.kr). (\textit{Corresponding Author: W. Shin})
}}

%
%


\maketitle

\begin{abstract}
The radio communication division of the International Telecommunication Union (ITU-R) has recently adopted Integrated Sensing and Communication (ISAC) among the key usage scenarios for IMT-2030/6G. ISAC is envisioned to play a vital role in the upcoming wireless generation standards. In this work, we bring together several paramount and innovative aspects of ISAC technology from a global 6G standardization perspective, including both industrial and academic progress. Specifically, this article provides 6G requirements and ISAC-enabled vision, including various aspects of 6G standardization, benefits of ISAC co-existence, and integration challenges. Moreover, we present key enabling  technologies, including intelligent metasurface-aided ISAC, as well as Orthogonal Time Frequency Space (OTFS) waveform design and interference management for ISAC. Finally, future aspects are discussed to open various research opportunities and challenges on the ISAC technology towards 6G wireless communications.
\end{abstract}


\IEEEpeerreviewmaketitle

\vspace{-3mm}

\section{Introduction}
\label{sec:1}

The ongoing standardization and implementation of Fifth Generation (5G) wireless networks have paved a shift towards exploring new technologies that can support the Sixth Generation (6G) wireless networks. A roadmap for a 6G terrestrial wireless network has been formed to deliver uninterrupted connectivity to both users and machine-type devices. For example, the radio communication division of the International Telecommunication Union (ITU-R) successfully drafted the new recommendation for the vision of International Mobile Telecommunication 2030 IMT-2030 (6G), which was recently approved at the meeting held in Geneva on June 2023.  As depicted in Fig. \ref{fig1a}, the development of IMT-2030 encompasses several emerging technology trends, including artificial intelligence (AI), Integrated Sensing and Communication (ISAC), sub-Tera Hertz (THz) transmission, channel adaption via reconfigurable intelligent surfaces (RIS) and holographic multiple-input multiple-output (MIMO) surfaces, etc. Specifically, ISAC possesses abilities to sense and better understand the physical world and transmission environment. 

ISAC is envisioned to play a key role in the upcoming generation. For example, integrated positioning, recognition, imaging, and reconstruction are expected to provide complimentary features that will be helpful in smart living, industrial advancements, and social governance. Moreover, the evolution of ISAC will further enhance wireless sensing capabilities and enable seamless collaboration between sensing and communication systems. Indeed, ISAC in 6G will unlock new possibilities for smart applications, enabling enhanced sensing capabilities, efficient resource utilization, etc. For example, some of the key aspects of ISAC in 6G are summarised below: 

\begin{figure}[t!]
    \begin{center}    
        \includegraphics[width=0.9\linewidth]{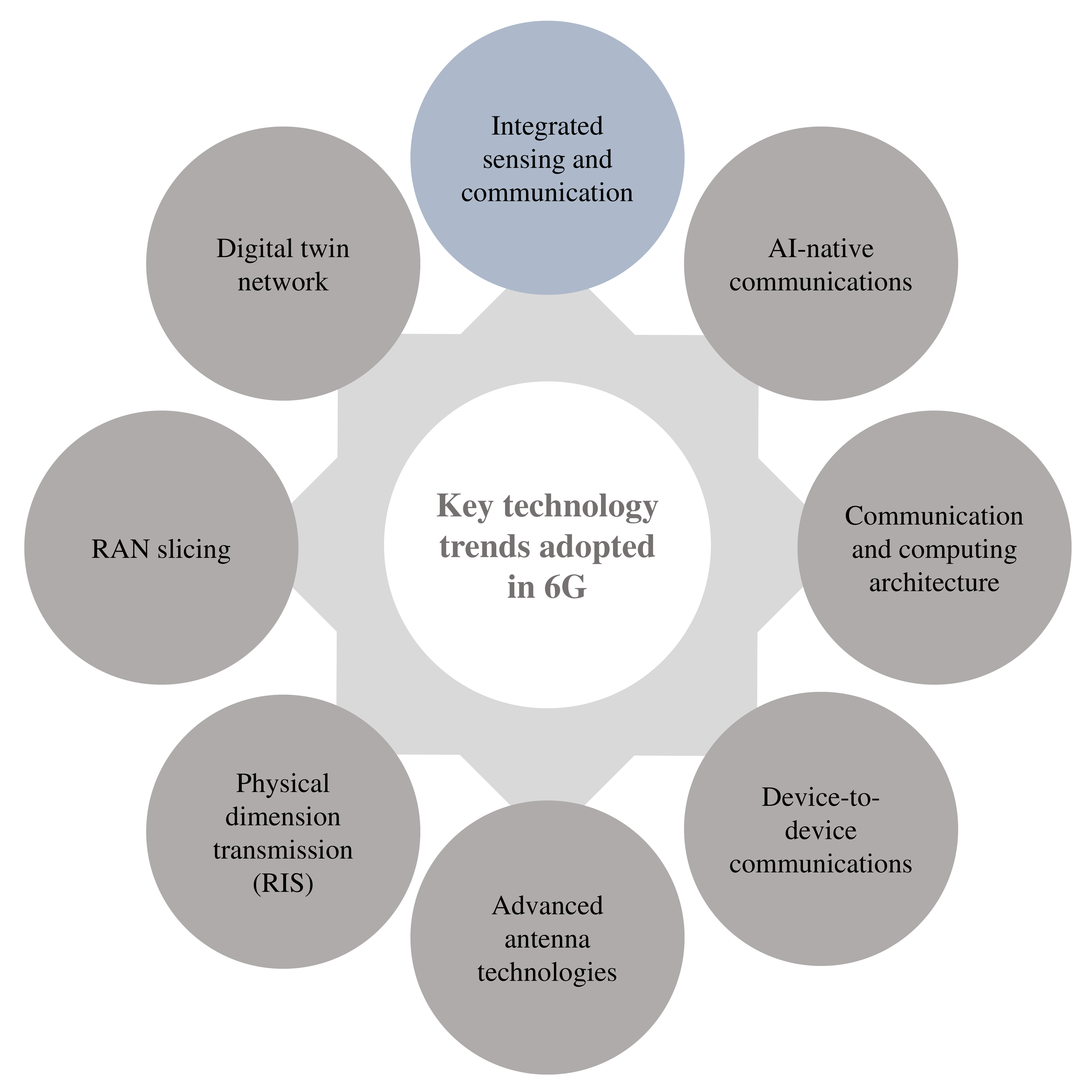}
    \end{center}
    \setlength{\belowcaptionskip}{-2pt}
    \caption{Key technology trends adopted in ITU-R FTT Report M.2516 \cite{itur}.}
    \vspace{-3mm}
    \label{fig1a}
\end{figure}

\begin{figure*}[t!]
    \begin{center} 
        \includegraphics[width=0.7\linewidth]{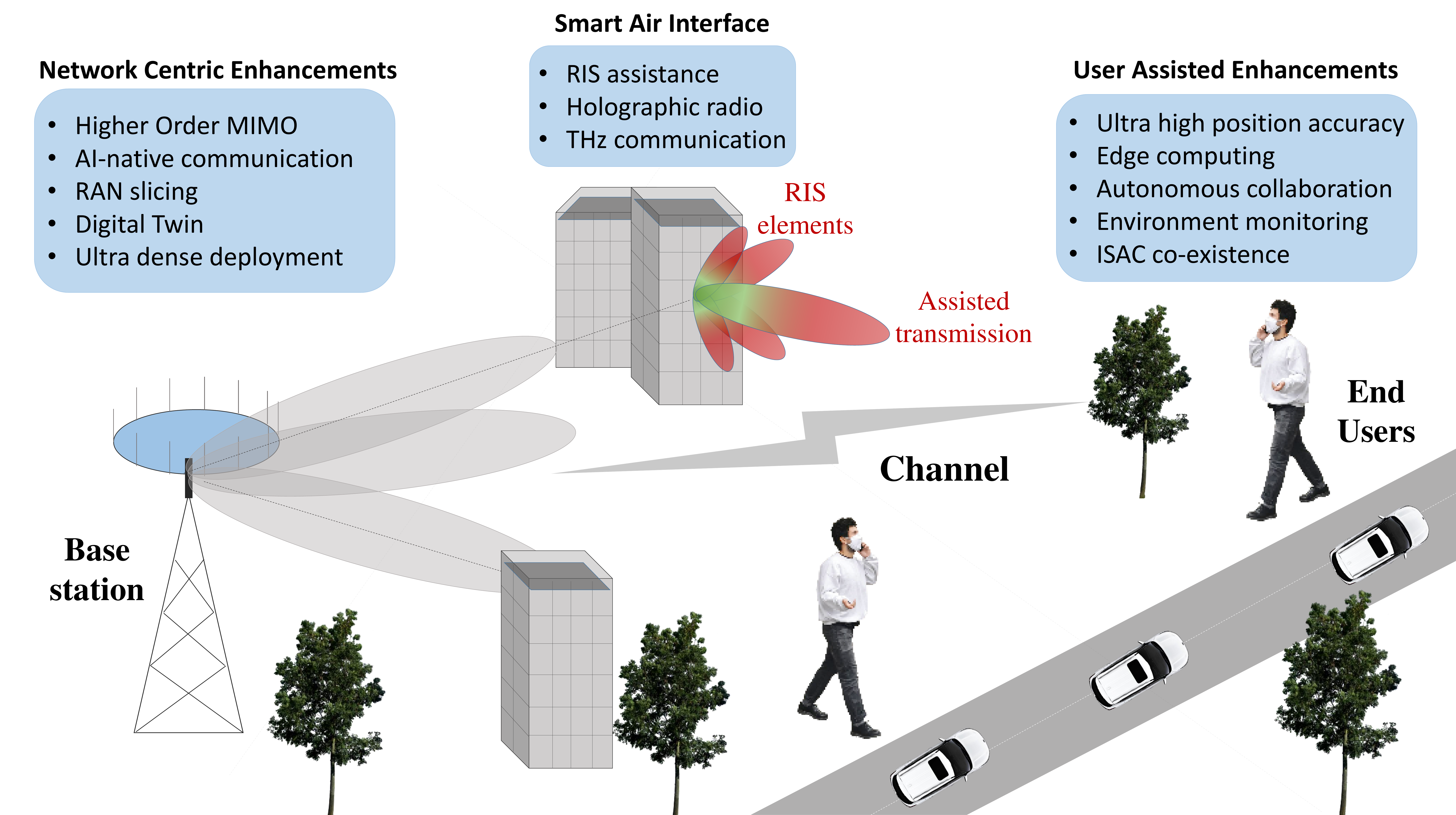}
    \end{center}
    \caption{An illustration of targeted 6G enhancements.}
    \label{fig2a}
    \vspace{-2mm}
\end{figure*}

\begin{itemize}

\item  \emph{Sensing-enabled Communication}: The utilization of integrated sensors enables real-time environmental monitoring, data collection, contextual awareness, etc. This, in turn, is expected to enhance the capabilities of wireless networks for intelligent information optimization, network management, etc. Moreover, ISAC can be used to assist wireless communication parameters, such as beamforming, channel allocation, etc \cite{l1}. 
\item  	\emph{Data fusion over Distributed Sensing}: A large number of sensors and devices are said to be facilitated under the 6G umbrella. Accordingly, the incorporation of ISAC towards the collection of data from various sources will allow more accurate and comprehensive transmission.
\item  	\emph{Multi-modal Sensing}: The co-existence of a diverse range of sensing modalities, including vision, audio, motion, environmental, biomedical sensing, etc., will provide information for applications such as augmented reality, immersive experiences, smart surveillance, healthcare monitoring, and environmental monitoring.
\item 	\emph{Intelligent Sensing and Communication Co-existence}: 6G standards are also expected to explore modern tools, including AI techniques for joint sensing and communication operations. Accordingly, AI-enabled ISAC will provide several benefits, including autonomous decision-making, intelligent data processing, intelligent vehicular networking, etc.
\end{itemize}

Understanding the need and emergence of ISAC in 6G, this article brings together several paramount and innovative aspects of ISAC for 6G, leading towards a paradigm shift in our current wireless standards. Specifically, this work highlights technical aspects and standardization of the upcoming wireless generation along with novel features and challenges and the latest industrial and academic progress on ISAC. Moreover, the paper summarises ISAC-enabled benefits from 6G purview, e.g., integrated localization, sensing, joint resource sharing, etc. Furthermore, the paper highlights several research directions,  opportunities, and use cases.      

\section{6G Overview: Technologies and Protocols}
\label{sec:2}
With the rapid pace of 5G implementation, the initiation of protocol formation for 6G technology is already underway. Indeed, the stepping stone has already been placed during the World Radiocommunication Conference 2023 (WRC-23). For ease of understanding, the proposed enhancements are grouped into three categories, as depicted in Fig. \ref{fig2a}.

\subsection{Network Centric Enhancements}
This encompasses both the evolution of current 5G capabilities and the incorporation of new techniques in 5G-Advanced and 6G. As the wireless framework's core, network-centric enhancements will be at the forefront of upcoming wireless generations. 

\begin{itemize}
    \item \emph{Evolution of Existing Capabilities:}  Experiences from the previous generations open new doors for the evolution of current capabilities. Leveraging advancements in spectrum efficiency, network capacity, etc., 6G aims to surpass the performance of its predecessors and unlock new possibilities. For example, a higher-order MIMO is intended to employ a larger number of antennas, enabling enhanced coverage and improved interference management. 

\item \textit{AI-Native Communication:} 6G is expected to embrace AI, enabling it to intelligently allocate resources, enhance network performance, improve energy efficiency, etc. AI-assisted communication is intended to leverage several new capabilities as an intrinsic component of the network. 

\item \textit{RAN Slicing:} Radio access network (RAN) slicing is one of the key techniques that enable the segregation of network-level responsibilities via slicing the network into multiple virtual networks, each tailored to specific use cases or service requirements. Accordingly, RAN slicing offers flexibility to support different industries, applications, etc. 

\item \textit{Digital Twin:} Digital twin is the modern learning tool to create a virtual replica of the physical network infrastructure. In the context of wireless communication, the notion of a digital twin provides real-time monitoring, simulation, and optimization capabilities. This, in turn, brings several innovative features, including the replication, update, and synchronization of the physical networks, etc.
\end{itemize}

\subsection{Smart Air Interface} 
Unlike previous wireless generations that employ processing at the transceiver ends, enabling 6G techniques will focus on the utilization of a smart air interface, making it convenient to tune the wireless channel for more favorable propagation conditions \cite{Marco}. Some of the potential techniques include RIS assistance, holographic radio, and THz communication. 

\begin{itemize}
    \item \textit{RIS Assistance:} RIS assistance is indeed an emerging component of 6G's smart air interface. The notion of RIS technology mainly involves the manipulation of wireless signals while reflected through passive intelligent surfaces \cite{rohit}. Enabled by channel estimation and dynamic phase tuning, RIS can help in several ways, e.g., assisting in beamforming, interference mitigation, coverage extension, etc. Already, potential use cases of RIS have been verified by numerous existing works.

\item \textit{Holographic Radio:} Holographic radio is another groundbreaking technique that comes under potential 6G enablers \cite{holo}. Specifically, it aims to transform several wireless aspects from signal transmission to reception and processing. Enabled by advanced signal processing algorithms and channel incoherence, holographic radio enables the simultaneous transmission of multiple data streams over the same time-frequency resource. 

\item \textit{THz Communication:} One step ahead, 6G frontiers are focusing on THz communication \cite{tera}, leveraging frequencies in the THz range (0.1-10 THz). Unlike lower frequency bands, THz waves offer wider bandwidths that in turn enable ultra-high data rates, potentially reaching multi-terabit-per-second speeds. Also, THz communication holds promise for data-hungry applications, e.g.,  holographic imaging, immersive virtual reality, etc.
\end{itemize}

\subsection{User Assisted Enhancements}
6G frontiers focus on utilizing the end user’s computational ability without significantly burdening the overall cost, size, and power consumption of the network system. 
\begin{itemize}
    \item \textit{Edge Computing:} Adding a nominal computational burden at the user’s end, several benefits could be achieved, including a) bringing computation and data storage closer to the network edge, b) reducing end-user latency, c) possessing real-time processing capabilities, and d) enabling context-aware services. Accordingly, 6G networks can support a diverse range of applications.

    \item \textit{Environment Monitoring:} By integrating sensors, devices, etc., into the edge-computing unit, wireless networks can provide real-time environmental insights to facilitate diverse applications, e.g., smart city initiatives, environmental sustainability, proactive decision-making, etc. Accordingly, this enables the networks to collect and analyze vast amounts of data for various modern applications.
    
    \item \textit{Ultra-High Position Accuracy:} Modern wireless transmission, including digital beamforming, interference management, etc., is pillared on the user’s positioning accuracy. 6G envisions ultra-high position accuracy by leveraging advanced positioning technologies, including a tight integration of communication and positioning. Ultra-high position accuracy opens up new 6G possibilities for applications like augmented reality, autonomous vehicles, and advanced logistics.
\end{itemize}

\section{6G Standardization and ISAC Co-existence}
\label{sec:3}

Standardization plays a crucial role in advancing ISAC by providing a common ground for developers, researchers, and industries to collaborate effectively. A comprehensive set of standards fosters compatibility and scalability and facilitates the integration of new technologies into existing ecosystems.

\subsection{Efforts and Organizations} 
\label{sec:3.1}

Embracing standardization will unlock the full potential of ISAC and pave the way for a connected, intelligent, and sensor-driven future. The major standard development organizations (SDOs) in the wireless communication domain, such as Third Generation Partnership Project (3GPP) and European Telecommunication Standard Institution (ETSI), will focus on innovative work on ISAC waveforms and processing for sensing. In the first preliminary phase, the Information Services Group (ISGs) will mostly define metrics for ISAC-relevant use cases for the trade-offs between resource allocation for communication and sensing. 

\subsection{\textcolor{black}{Third Generation Partnership Project (3GPP)}}
\label{sec:3.1}
An interest in ISAC is gradually increasing after the first stage of the 5G standard as the scope of 5G-Advanced. While the conventional 5G providing only communication services was the core of its function, ISAC can be important in its function to contribute to communication-assisted new services. To identify such new ISAC service scenarios, three key scenarios are discussed: object detection and tracking, environment monitoring, and motion monitoring.
\begin{itemize}
    \item \emph{Object Detection and Tracking:} When user equipment (UE) is connected to a network, various methods such as positioning and UE’s feedback make the network track the UE. However, the network cannot obtain such information from the passive object or targets without having the network connection functionality. In this scenario, high accuracy of tracking positions and velocities will be the key to tracking both connected and unconnected devices.
    \item \emph{Environment Monitoring:} In this scenario, the base station is used as an environmental sensor using the existing communication equipment. In particular, a sensing device will obtain environmental information such as the existence of rainfall, the level of flood, human gathering, and traffic-load information. Since monitoring information is time-critical, sensing latency of 1 minute or less is necessary.
    \item \emph{Motion Monitoring:} Indoor human motion, sleep monitoring, sports monitoring, and gesture recognition are use cases in this scenario. It is very crucial to minimize false detection rather than location accuracy.
\end{itemize}

In 3GPP, the second phase of the study will continue to identify the ISAC-related key requirements further. At the same time, RAN, which is responsible for radio access technologies, needs to study whether physical layer support for ISAC is necessary for 5G-Advanced. However, existing channel models only defined the various characteristics of multi-paths between transmitters and receivers in a stochastic approach. In the case of sensing, it is necessary to consider the reflectors of explicit surrounding objects and buildings and how a certain cluster contributes either to both communication and sensing. The channel model study in which these characteristics are considered will be the beginning of physical layer (PHY) design of ISAC. 

\subsection{\textcolor{black}{European Telecommunications Standards Institute (ETSI) }}
\label{sec:3.2}
Hereafter, we discuss the significance of standardization in addressing interoperability challenges, ensuring data integrity, and mitigating potential risks associated with ISAC deployment from a European standardization perspective. In particular, the ETSI has been at the forefront of standardization activities related to communication and networking technologies. With the emergence of ISAC, ETSI's role becomes even more vital in providing a conducive environment for stakeholders to collaborate and develop unified specifications.
According to the ETSI Technology Radar (ETR), ISAC-oriented standardization activities will focus on the following aspects:
\begin{itemize}
    \item ISAC waveform, sequence, coding, modulation, and beamforming;
    \item MIMO, massive MIMO, and RIS for ISAC;
    \item Passive sensing using communication waveforms;
    \item Millimeter wave (mmWave) and THz ISAC.
\end{itemize}

As ISAC encompasses a wide range of technologies, harmonizing standards can be complex, particularly considering the diverse application domains and varying requirements. Moreover, the rapid pace of technological advancements necessitates adaptive and forward-looking standardization approaches. This would have a direct liaison with other running ETSI groups, focusing on emerging technologies, such as ETSI RIS, on RIS topics, and ETSI THz, on THz communications.

\section{Intelligent Metasurfaces-aided ISAC and Novel ISAC Waveform Design}

\subsection{RIS and Holographic MIMO-assisted ISAC}
The integration between wireless communication networks and sensing systems has already been investigated for some years. Mainly the focus has been on enabling the co-existence between the two systems on the same frequency spectrum rather than implementing a true integration of the two functions in a shared platform. Instead, the vision of ISAC in 6G is to obtain a tight integration of the two functionalities, in which a single hardware platform transmits a single waveform that is designed to carry information symbols as well as perform environment sensing at the same time. In principle, this could be achieved by adopting large multiple antenna arrays, e.g., massive MIMO, but practical considerations prevent from taking this approach. In other words, while deploying massive antenna arrays provides higher rates, it also leads to a significant energy consumption increase \cite{eMIMO}. In order to successfully integrate communication and sensing in a single platform, the required radio resources must be secured with sustainable energy consumption. One approach with the potential to achieve this goal is based on the recent technological breakthrough of holographic MIMO beamforming. The main idea behind this approach is to equip wireless transceivers with reconfigurable metasurfaces, which can be used for transmit and/or receive beamforming with much lower energy requirements. 

Intelligent metasurfaces provide a novel approach for wireless communications, in which planar structures of metamaterials are equipped with elementary electromagnetic (EM) units spaced at sub-wavelength distances with each other. Each unit can apply an individual EM response to an input signal, operating directly in the analog domain, i.e., without the need to employ a digital transceiver chain.  The use of this novel transceiver architecture allows one to feed $L$ data streams to $K$ radio frequency chain, with $K\ll L$ in a similar way in which hybrid MIMO architectures work. After impinging on the metasurface, the signal travels to the receiver, where another metasurface can be placed in the field of the receive radio-frequency chains. This architecture has two key advantages compared to traditional multiple-antenna architectures and hybrid MIMO architectures:
\begin{itemize}
\item Compared to existing hybrid MIMO architectures, it provides a much larger number of free parameters that can be optimized for the maximization of the system performance. Moreover, the reflection coefficients can be reconfigured in real-time, in order to adapt to fluctuations and variations of the electromagnetic channel between transmitter and receiver. 
\item Compared to existing massive MIMO architectures, the large number of free parameters for system design comes at lower energy consumption and cost. Reconfigurable metasurfaces can be nearly-passive devices, only requiring a small amount of energy to power the hardware components that enable the reconfiguration (low-power switches like PIN diodes or varactors). Moreover, the metasurfaces operate in the analog domain without requiring energy-consuming conversion to/from the digital domain. 
\end{itemize}

In other words, holographic MIMO beamforming by a metasurface can be seen as a planar massive MIMO array with much lower energy requirements and cost and a large amount of radio resources that can be optimized for optimal system performance. Having many free parameters for signal optimization appears to be a critical requirement for the successful integration of communication and sensing, not only in order to synthesize suitable waveforms but also to balance between the different metrics that should be optimized. In an ISAC system, multiple, possibly contrasting, performance metrics must be optimized, e.g., the communication rate, latency, detection rate, estimation accuracy, energy efficiency, etc. Besides beamforming, another crucial aspect is the efficient ISAC waveform design presented in the following. 

\subsection{OTFS Waveform Design for ISAC}
6G and beyond wireless systems are likely to incorporate higher frequency bands, such as mmWave and THz. Besides, these systems are expected to support  high mobility communications with speeds greater than 1,000 km/h, such as hyper-high-speed railway (hyper-HSR) and airline systems \cite{2921208}. As a result, future wireless systems will experience doubly dispersive channels that vary in both time and frequency. However, the performance of the classical orthogonal frequency division multiplexing (OFDM) modulation suffers from high Doppler spread, specifically in very high frequencies where the channel is doubly dispersive. This limits the suitability of OFDM in future 6G networks. This has introduced a requirement for new communication technologies capable of handling high Doppler spread introduced by the doubly dispersive channels. 

Motivated by the deterministic nature of the wireless channel in the delay-Doppler domain (DD-domain), the information symbols of the orthogonal time-frequency space (OTFS) waveforms are modulated in the DD-domain instead of the conventional time-frequency (TF) domain. The use of OTFS modulation in communication systems is well studied, and it is shown that the communication error performance of OTFS can be improved further by use of low-density parity-check (LDPC) and convolution-coded OTFS \cite{3071493}.

\begin{figure*}
     \centering
     \begin{subfigure}[t]{0.45\textwidth}
         \centering
         \includegraphics[width=\textwidth]{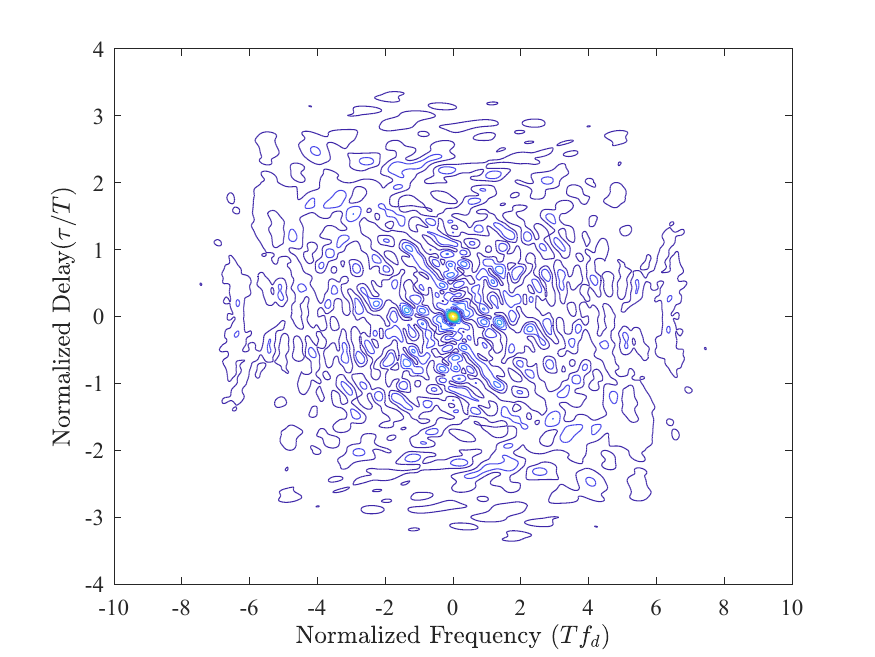}
         \caption{First realization of communication data.}
     \end{subfigure}
     \hfill
     \begin{subfigure}[t]{0.45\textwidth}
         \centering
         \includegraphics[width=\textwidth]{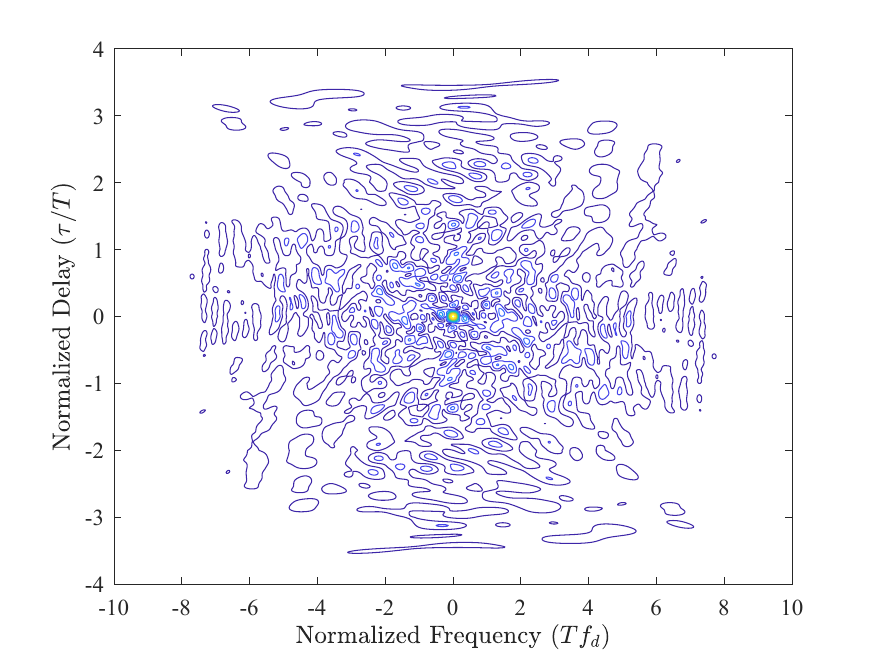}
         \caption{Second realization of communication data.}
     \end{subfigure}
     \hfill
        \caption{The Ambiguity Function (AF) of modulated ISAC-OTFS waveform.}
        \label{figure_OTFS_1}
        \vspace{-3mm}
\end{figure*}

\begin{figure}
    \centering
    \includegraphics[width=0.45\textwidth]{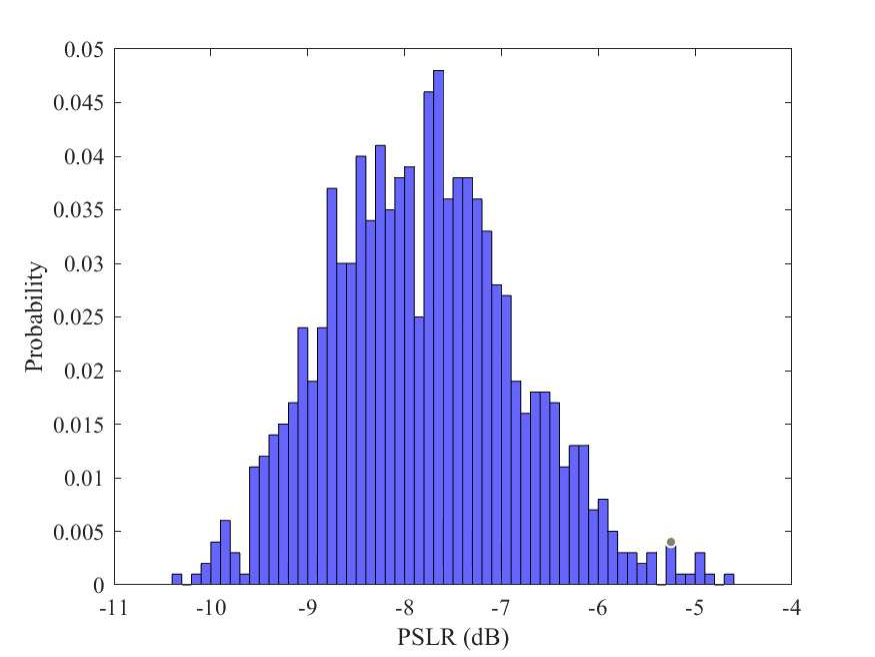}
    \captionsetup{justification=centering}
    \caption{The PSLR distribution of modulated ISAC-OTFS waveform.}
    \label{figure_OTFS_2}
    \vspace{-4mm}
\end{figure}

Coming to the ISAC counterpart, the modulation of communication data changes the properties of the OTFS waveform used in radar sensing. As such, it affects the performance of radar sensing. To further illustrate this, let us consider the ambiguity function that represents the response of a matched filter when the transmitted signal is received with a certain time delay and a Doppler shift. As a simple illustration, Fig.~\ref{figure_OTFS_1} plots the contour plot of the ambiguity function for two realizations of $4$-quadrature amplitude modulation (QAM) modulated ISAC-OTFS waveform with different communication data matrices of size $4\times 8$ in DD-domain. From Fig.~\ref{figure_OTFS_1} we can observe that the behavior of the ambiguity function, especially outside the mainlobe regime, depends on the communication data modulated onto the OTFS waveform. Indeed, recent research has shown that both OTFS and OFDM-based ISAC systems provide as accurate local radar estimates (delay and Doppler resolution and mean-squared error of estimates) as frequency-modulated continuous wave (FMCW) while OTFS-based systems provide a higher pragmatic capacity for communication \cite{2998583}. Nevertheless, the global accuracy in radar sensing has received limited attention and a detailed analysis of the impact of data modulation on global radar performance is essential for the successful implementation of OTFS modulation in future ISAC systems \cite{3209651}.

Furthermore, the peak-to-sidelobe ratio (PSLR) is a well-known global radar performance metric used to identify the capability of detecting weak targets in the presence of nearby interfering targets \cite{7507195}. As a simple illustration, Fig.~\ref{figure_OTFS_2} plots the distribution of the PSLR for $4$-QAM modulated ISAC-OTFS waveform with $1,000$ random communication data matrices of size $4\times 4$ in DD-domain. From Fig.~\ref{figure_OTFS_2}, we can clearly observe that the PSLR of OTFS varies from $-10.3$ dB to $-4.6$ dB based on the modulated communication data. Due to the smaller probability of false alarms, a smaller PSLR value is a desired property in ISAC systems.

\subsection{ISAC Interference Management}

\begin{figure}[t!]
    \centering
    \includegraphics[width=0.49\textwidth]{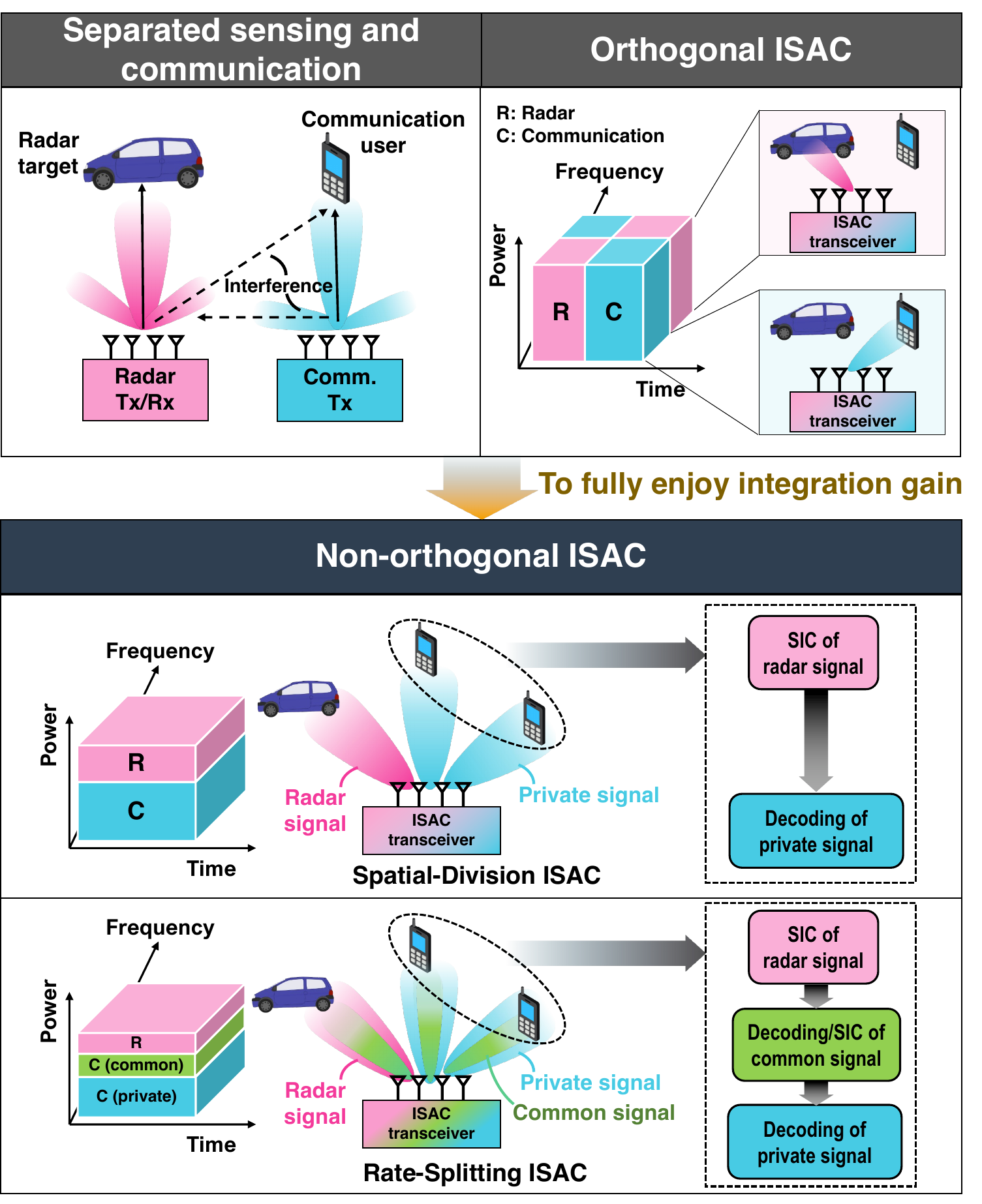}
    \caption{The evolution of ISAC from orthogonal to non-orthogonal approaches for efficient use of wireless resources.}
    \label{Fig_Interference_management}
\end{figure}

Furthermore, ISAC has great potential for more efficient resource utilization by integrating communication and sensing into a single system, which has traditionally been performed on separate hardware. In addition to the existing intra-system interference within the communication and sensing systems, the co-existence of sensing and communication within a single system brings out a new challenge in managing inter-system interference between the two functionalities. Two candidate approaches exist for tackling this problem: orthogonal ISAC, where time or frequency resources are allocated orthogonally, and non-orthogonal ISAC, which has the opportunity to maximize resource efficiency by sharing time and frequency resources for both functions.
\subsubsection{Orthogonal ISAC}
As shown in Fig. \ref{Fig_Interference_management}, the orthogonal allocation of time or frequency resources provides a straightforward way to manage interference between the two functions.  In time-division ISAC, different waveforms can be utilized for communication and sensing in separate time slots. On the other hand, frequency-division ISAC typically employs OFDM waveform, with the allocation of distinct subcarriers for different functions to achieve interference mitigation \cite{liu2022integrated}.

\subsubsection{Non-orthogonal ISAC}
The non-orthogonal ISAC has attracted a great deal of interest due to its ability to take full advantage of resource efficiency. This approach can be broadly categorized as follows:
\begin{itemize}
    \item \textit{Spatial-division ISAC:}
    By spatially dividing beams for communication and sensing purposes, spatial-division ISAC has the potential to achieve superior spectral efficiency, compared to orthogonal ISAC. With a sufficient number of antennas to form a narrow beam, it is effective when line-of-sight (LOS) paths dominate the communication channel.
    However, severe interference can occur in scenarios where the target is near multiple users or in a rich scattering environment. To address this issue, introducing additional radar signals and corresponding successive interference cancellation (SIC) can be a solution, as shown in Fig. \ref{Fig_Interference_management}.
    \item \textit{Rate-splitting ISAC:} Rate-splitting ISAC is based on rate-splitting multiple access (RSMA), gaining attention as the next generation of multiple access schemes. In RSMA, messages intended for each user are split into common and private signals and transmitted. 
    The {common signal} plays a triple role in rate-splitting ISAC: intra-system interference management, inter-system interference management, and beam forming to the target. Indeed, it has been studied that the {common signal} can function like a radar signal regardless of whether an additional radar signal is employed. In addition, there is no loss of performance in the absence of a radar signal thanks to the common stream in RSMA \cite{xu2021rate}.
\end{itemize}

Further research is essential to mitigate such interference by exploring advanced user scheduling, frame structure design, and cooperative strategies using multiple ISAC transceivers, etc. Addressing these issues will play a key role in improving the performance of ISAC systems.

\section{ISAC Challenges and Considerations}
\label{sec:4}

Formally, 6G has been envisioned as the successor to the previous wireless generation with greater capabilities. However, it is not merely an evolution of its predecessors but rather a paradigm shift to enable novel use cases that would be difficult to support over the precedent wireless generations. Thereby, to fulfill its ambitious vision, there are various critical challenges that need to be resolved.

\begin{itemize}
    \item \textit{Inter-operability Harmony:} Inter-operability harmony is essential for the seamless integration of sensing and communication technologies, enabling devices to function as sensors and communication nodes. However, it fosters a collaborative environment which is a challenging task. To achieve inter-operability harmony among communication and sensing units, collaborative efforts between industry stakeholders and standardization bodies are required to establish common protocols. Moreover, other integration challenges include resolving interoperability issues across heterogeneous networks and ensuring data privacy and security concerns. Overall, inter-operability harmony is a fundamental aspect of ISAC to unlock new possibilities for innovative applications and drive the next generation of intelligent connectivity.

\item \textit{Security and Privacy:} Another big challenge arises due to the breaching of data secrecy, transmitted on the top of sensing signals. In addition to the numerous possibilities, the \emph{comm-sense} fusion of ISAC also brings forth significant security and privacy concerns. ISAC networks superimpose sensor data and communication that requires robust security measures to protect against cyber threats, data breaches, and unauthorized access. Ensuring data privacy is essential to build trust among users to safeguard personal information and sensitive data from unauthorized access or misuse. Collaboration among industry stakeholders, researchers, and regulatory bodies is necessary to establish common security standards, best practices, and guidelines. The security and privacy of ISAC networks are of paramount importance to ensure trust, reliability, and widespread adoption of intelligent connectivity.\cite{sec}
\end{itemize}

\section{ISAC Future Prospects}
\label{sec:5}
Being key enabling technology for 6G, ISAC is at the forefront of wireless communication, which opens several doors for research directions and other opportunities. 

\begin{itemize}
    \item \textit{Emerging Trends in ISAC:} Several trends are shaping the evolution of ISAC networks: \textit{a)}	ISAC-enabled transmission will be increasingly harnessing edge computing capabilities to process data closer to the source, providing several benefits, including reduced latency, optimized bandwidth, and enhanced real-time data analytics, \textit{b)} AI and machine learning will play a vital role in ISAC networks, enabling advanced data processing, intelligent resource allocation, and self-optimizing network behavior, \textit{c)}
ISAC networks will enable collaborative communication, where devices and sensors work together to share information and collectively improve system performance. 

\item \textit{Envisioned Use Cases and Applications:} ISAC co-existence is set to revolutionize the landscape of forthcoming wireless communication, promising several new opportunities including: \textit{a)} ISAC is intended to play a pivotal role in shaping smart cities of the future, enabling real-time monitoring of traffic flow, energy distribution, public safety (such as response and recovery in a disaster-affected region), etc. By integrating sensors with communication capabilities, smart cities can optimize several things, e.g., resource allocation, reduce congestion, etc, \textit{b)} The fusion of sensor data and communication makes ISAC the core of Industry 4.0/5.0, revolutionizing industrial automation. ISAC is set to provide real-time monitoring of machines and processes, facilitating predictive maintenance, etc, and \textit{c)} ISAC is set to drive transformative applications, especially in the realm of autonomous vehicles. Moreover, integrated sensors and communication enable real-time data sharing among vehicles, making it convenient to offer improved safety, navigation, and traffic management. 
\end{itemize}

Undoubtedly, ISAC technology holds immense promise in transforming households and industries. Countless envisioned use cases and applications of diverse nature, including smart cities, healthcare, industrial automation, environmental monitoring, agriculture, transportation, and many more. 

\section{Conclusion}
{Understanding the importance and emergence of ISAC in the upcoming wireless generation, we have presented several essential and innovative aspects of ISAC technology from a 6G standardization purview. Specifically, this work can be concluded as: \textit{a)} This article summariz	es 6G requirements and the vision of ISAC integration, covering various aspects of 6G standardization, advantages of ISAC co-existence, and related challenges, \textit{b)} Additionally, the article has highlighted key enabling technologies, e.g., intelligent metasurface-aided ISAC, and also presents the OTFS waveform design for ISAC, \textit{c)} Moreover, the article has explored future possibilities, opening the doors for various research avenues concerning ISAC technology for 6G wireless communication.}


\bibliographystyle{IEEEtran}

\bibliography{IEEEabrv,BibRef}


\end{document}